# A Cross-Corpus Speech Emotion Recognition Method Based on Supervised Contrastive Learning

*Minjie Xiang*

School of Information Science and Engineering, Southeast University; Nanjing China; 210089

**Abstract**: Research on Speech Emotion Recognition (SER) often faces challenges such as the lack of large-scale public datasets and limited generalization capability when dealing with data from different distributions. To solve this problem, this paper proposes a cross-corpus speech emotion recognition method based on supervised contrast learning. The method employs a two-stage fine-tuning process: first, the self-supervised speech representation model is fine-tuned using supervised contrastive learning on multiple speech emotion datasets; then, the classifier is fine-tuned on the target dataset. The experimental results show that the WavLM-based model achieved unweighted accuracy (UA) of 77.41% on the IEMOCAP dataset and 96.49% on the CASIA dataset, outperforming the state-of-the-art results on the two datasets.

**Key words**: Speech emotion recognition; Contrastive learning; Self-supervised features

## 1  Introduction

Recognition of emotion in speech is a key technology in human-computer interaction. With the increasing application of dialogue systems, the demand for SER tasks is also increasing [1]. General deep learning models often require a large amount of data to achieve good results and strong generalization ability, while in the field of SER, there is a lack of large-scale public datasets, and SER tasks also face the problem of weak generalization ability due to the gap between different languages and speakers [2]. The emergence of self-supervised learning speech representation models, which are often trained on hundreds of hours of speech recognition datasets and can be used for numerous downstream tasks related to speech, provides a viable solution to these problems. Many studies [3][4][5] design fine-tuning algorithms for SER tasks based on pre-trained speech representation models. But these studies tend to finetune only a single dataset, ignoring the gains in model performance from using information from SER datasets in different languages.

Contrast learning has been used in self-supervised learning in image field in recent years. Positive and negative pairs are important concepts in contrast learning. The goal of contrast learning is to minimize the gap between positive pairs and maximize the gap between negative pairs. In order to make effective use of the data, this paper designs a supervised contrast learning algorithm to further fine-tune the speech representation model, taking speech samples from the same or different languages but with the same emotion as a set of positive pairs, and samples from different languages or different emotions as negative pairs, optimizing the speech representation model by minimizing contrast loss and cosine interval loss. After fine-tuning the speech representation model, a second fine-tuning is performed on different datasets to obtain the classifier. In order to simplify the experiment, only two SER datasets are considered in this paper, namely the English SER dataset IEMOCAP[6] and the Chinese SER dataset CASIA[7]. The experimental results show that the UA of the model based on the speech representation model

Hubert[8] on the IEMOCAP dataset and the CASIA dataset are 76.92% and 97.19%, respectively, and the UA of the classification model based on WavLM[9] on the two datasets are 77.41% and 96.49%, respectively.

The follow-up arrangement of this paper is as follows: the second chapter introduces the work related to SER task of fine-tuning speech representation model and contrast learning, the third chapter discusses the proposed algorithm in detail, the fourth chapter introduces the experimental setting and analyzes the experimental results, and the fifth chapter summarizes the research results and briefly looks forward to the future research direction.

## 2  Related Work

In recent years, many studies have applied self-supervised learning to speech representation, and speech representation models such as Wav2Vec2.0[10], Hubert and WavLM have successively appeared, and these models are often fine-tuned for SER tasks. The simplest method of fine-tuning is to perform supervised training on the SER dataset after freezing some parameters. The features of all time steps need to be aggregated before classification, and many studies use different aggregation methods when fine-tuning. For example, Morais et al. [11] used ECAPA-TDNN[12], which was originally used to extract speech embedding features, to aggregate features, and Kakouros et al. [13] used Attentive Correlation Pooling (ACP) to aggregate time step features. Some studies have proposed new fine-tuning algorithms. For example, Chen et al. [3] used Pseudo-label task adaptive pretraining (P-TAPT) to fine-tune speech representation models.

Contrast learning is based on the simple idea of learning deep features of objects by asking the model to compare the similarities between similar objects (positive pairs) and the differences between different objects (negative pairs). In the field of speech, in addition to generating positive pairs using image-like enhancement methods, data from different time steps can also be used as positive and negative pairs. For example, Li et al. [14] used contrast prediction coding to perform SER, allowing the model to predict the data of several future time steps only through the current and past information, taking the real data of several future time steps as a positive pair and the randomly sampled data as a negative pair. Ulgen et al. [5] assumed that emotional information exists in speaker features, and used k-means to cluster speaker features. Positive pairs were samples sampled in the same cluster for the same speaker, while negative pairs were samples sampled in different clusters for the same speaker. Considering the continuity of phonetic emotion, Xu et al. [15] took the emotion of the current pronunciation and the previous pronunciation of the same speaker in a dialogue as positive pairs, and the other pairs as negative pairs. In order to extract features from speech, many contrast learning methods also use pre-trained speech representation models.

## 3  Methods

3.1 Two-stage fine tuning

The two-stage fine-tuning framework is shown in Figure 1. The same basic architecture is used in the implementation of the speech representation model. The original speech is first encoded into high-dimensional feature representation by a multi-layer convolution-based speech encoder, and then dimension is enhanced by a feature projection built at the full connection layer, and features are further extracted by the feature encoder of the Transformer architecture. In the first stage of fine tuning, the parameters of the speech encoder and the feature projection are fixed, and only the parameters of the feature encoder are optimized. The self-attention pooling layer refers to the design in Safari et al. [16], which uses a simple attention mechanism to aggregate the features of all time steps, assuming that the output of the feature encoder is $H = [h_1, h_2, \cdots, h_T]^T \in \mathbb{R}^{T \times d}$, The output $C$ of the self-attention pooling layer is

$$C = \text{softmax}(W_c H^T) H \quad (1)$$

Where $W_c \in \mathbb{R}^d$ is trainable parameters. The structure of the classifier is two fully connected layers, with the ReLU activation function in the middle. The classifier is only used when fine-tuning the classifier in the second stage, and the loss function in the second stage is cross entropy loss.

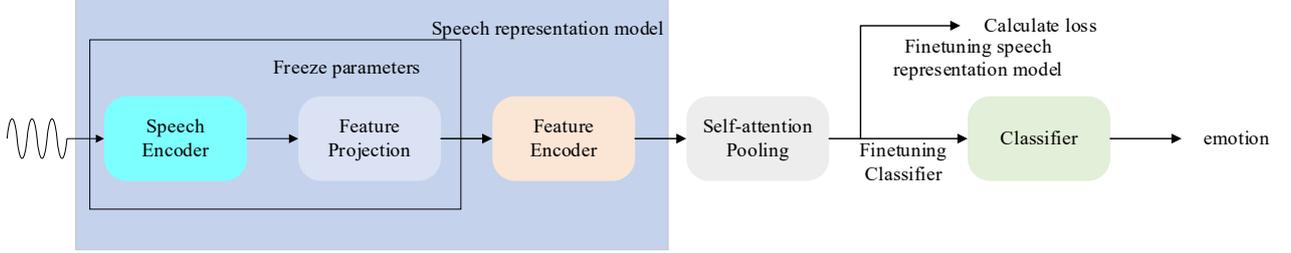

Figure 1 Two-stage fine-tuning frame

In the first stage of fine-tuning, the supervised contrast learning method was used to optimize the model. Samples of one batch were sampled from two datasets in each iteration. Assuming the batch size was $N$, the process of sampling positive and negative pairs was shown in figure 2 and figure 3. Firstly, an emotion category $e$ is randomly determined from all the emotion categories as the emotion category when extracting the positive emotion category. If emotion category $e$ exists in both datasets, then $N/4$ samples with $e$ are randomly selected from each dataset. If $e$ appears in only one dataset, then randomly sample $N/2$ samples of $e$ from this dataset, for a total of $N/2$ positive samples. Negative samples are randomly selected from two datasets respectively, $N/4$ samples whose emotion category does not belong to $e$, and a total of $N/2$ negative samples are collected.

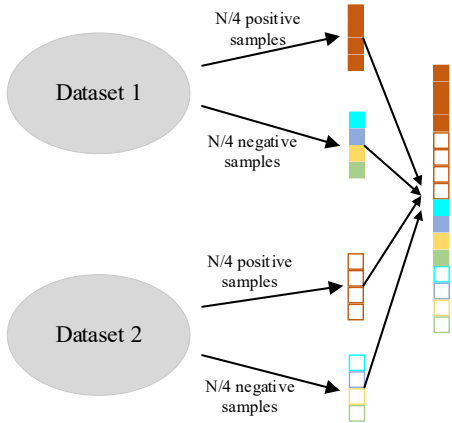
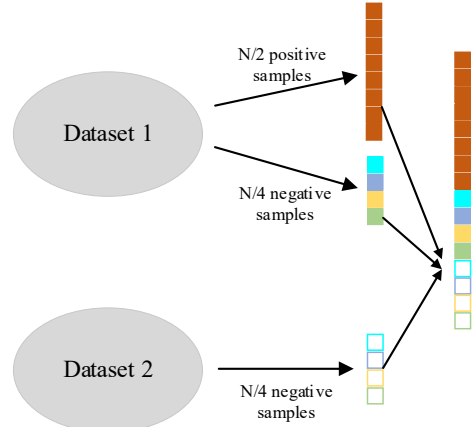

Figure 2 Sampling process of two datasets has $e$     Figure 3 Sampling process when only dataset 1 has $e$

2.2 Contrast loss

The contrast loss is InfoNCE loss, suppose $x$ is the feature obtained by aggregation of the self-attention pooling layer, $x_p^i$ is the feature of the i-th positive sample, and $x_n^i$ is the feature of the k-th negative sample. One positive sample in a batch forms a positive pair with other positive samples and a negative pair with all negative samples. The formula for calculating the contrast loss $L_c$ is

$$l_{ij} = \frac{\exp(\text{sim}(x_p^i, x_p^j)/\tau)}{\exp(\text{sim}(x_p^i, x_p^j)/\tau) + \sum_{k=1}^{N/2} \exp(\text{sim}(x_p^i, x_n^k)/\tau)} \quad (2)$$

$$L_c = \frac{4}{N(N-2)} \sum_{i=1}^{N/2} \sum_{j=i+1}^{N/2} \left( l_{ij} + l_{ji} \right) \quad (3)$$

Where $\mathrm{sim}(x, y) = x^T y / \|x\| \cdot \|y\|$, $\tau$ is the hyperparameter, which is 0.07.

2.3 Cosine margin loss

To further narrow the distance between similar samples and avoid the difference between different samples, cosine margin loss is introduced. When calculating the loss, the positive sample equal quantity is split into two parts, called x and y. The loss formula is as follows

$$L_m = \alpha \left( 1 - \mathrm{sim}\left( x_{p_1}, x_{p_2} \right) \right) + \max \left( 0, \mathrm{sim}\left( x_p, x_n \right) - m \right) \quad (4)$$

Where $\alpha$ and $m$ are hyperparameters, $\alpha$ is 0.5 and $m$ is 0.4. Finally, the total model loss is the sum of contrast loss and cosine margin loss.

$$L = L_c + L_m \quad (5)$$

## 4 Experiment and Result Analysis

4.1 Datasets

The two datasets used in this experiment are IEMOCAP and CASIA. IEMOCAP is a common dataset used for SER tasks, and because of the uneven distribution of emotions across the various categories of the IEMOCAP dataset, only four of them are considered: neutral, sad, angry, and happy. CASIA is one of the most commonly used datasets in Chinese speech emotion recognition, containing a total of six emotions, which cover the four emotions in the IEMOCAP dataset , in addition to the two emotions of fear and surprise.

4.2 Experiment settings

The speech representation model Hubert[1] and WavLM[2] used in this experiment are both base versions downloaded from huggingface, and the number of parameters is about 95M. When fine-tuning the speech representation model, the pre-trained model parameters are loaded first, the parameters of the speech encoder and feature projection are fixed, and only the parameters of the feature encoder are optimized. The initial learning rate for the first stage of fine tuning is 1e-4, using the Adam optimizer. A total of 50 epochs are trained, starting from the 25th epoch, the learning rate is reduced by half every 5 epochs, the learning rate of the feature encoder is 0.4 times the learning rate, and the batch size is 32. When fine-tuning the classifier in the second stage, the parameters of the speech representation model and the self-attention pooling layer will be loaded, and the parameters of the whole speech representation model will be fixed at the same time, and only the parameters of the self-attention pooling layer and the classifier will be fine-tuned. The learning rate is 5e-4. Adam optimizer will be used to train 10 epochs, and the learning rate will be reduced to 0.1 times of the original after 5 epochs. The batch size is 32. The output dimension of the encoder is 768, and the output dimension of the two fully connected layers in the classifier is 256 and the number of affective categories.

---

[1] https://huggingface.co/facebook/hubert-base-ls960
[2] https://huggingface.co/microsoft/wavlm-base-plus

In order to evaluate the performance of supervised contrast learning, an experiment was also conducted to directly fine-tune the speech representation model (FT). During the experiment, the initial learning rate was 1e-3, and the Adam optimizer was used to train 50 epochs in total. The learning rate became 0.2 times of the original every 20 epochs, and the batch size was 32. For the speech representation model, the parameters of the speech encoder and the feature projection are also fixed, and the parameters of the feature encoder are optimized with only 0.4 times the learning rate.

During the training, the 5-fold cross-validation method was adopted, and the two datasets were divided into 5 parts respectively, leaving one part for testing. In the first stage of fine-tuning, only the data used for training was sampled. The final evaluation of the model is the average of the 5 experimental results.

3.3 Compared with methods for fine-tuning speech representation models

Table 1 shows the performance comparison between the supervised contrast learning fine-tuning method proposed in this paper and other fine-tuning methods in the IEMOCAP dataset. Input features of all models are original speech, excluding multimodal features such as text and video.

Table 1 Unweighted accuracy UA (%) on the IEMOCAP dataset

| # | Speech representation model | Unweighted accuracy (UA) |
|---|---|---|
| 1 | Hubert-base(FT) | 66.61 |
| 2 | WavLM-base(FT) | 68.31 |
| 3 | Wav2vec2.0(P-TAPT)[3] | 74.30 |
| 4 | Hubert(ACP)[13] | 73.86 |
| 5 | WavLM(ACP)[13] | 75.60 |
| 6 | Hubert(ECAPA-TDNN)[11] | 76.78 |
| 7 | Wav2vec2[5]* | 73.80 |
| 8 | Hubert-base* | 76.92 |
| 9 | WavLM-base* | **77.41** |

* Indicates that comparative learning is used

Comparing experiments 1 and 2 with experiments 8 and 9, it can be seen that the supervised contrast learning proposed in this paper can effectively improve the classification performance of the model compared with direct fine tuning, which increases by 10.31% and 9.1% respectively. Compared with other fine-tuning methods (experiments 3, 4, 5, 6), there is also a relatively obvious improvement. Compared with experiments 4 and 6, it can be seen that different polymerization methods have a considerable impact on model performance, and more complex aggregators can more effectively characterize features. Compared with speaker characteristics in experiment 7, speech contrast learning using the same emotion of the two datasets can help the model ignore the interference of emotion-irrelevant factors in speech to a certain extent and pay attention to the internal representation of emotion.

3.4 Compared with other conventional speech emotion recognition models

Table 2 shows the performance compared to other conventional SER models, which are models that do not use pre-trained speech representation but are designed for SER tasks. It can be seen from the table that the fine-tuned speech representation model on IEMOCAP and CASIA datasets has performance advantages over other specially designed SER models on both datasets, which proves that the proposed method can effectively learn the information of both datasets at the same time. The fine-tuned speech representation model is more generalized by using information from other datasets.

Table 2 Unweighted accuracy UA (%) on IEMOCAP and CASIA datasets

| Model | IEMOCAP | CASIA |
| --- | --- | --- |
| Dual-TBNet[17] | 66.70 | 95.70 |
| TIM-Net[18] | 72.50 | 94.67 |
| EmoDARTS[19] | 76.56 | — |
| Hubert-base | 76.92 | **97.19** |
| WavLM-base | **77.41** | 96.49 |

3.5 Feature similarity analysis

Figure 4 shows the original WavLM (left) and fine-tuned WavLM (right) of the positive and negative samples respectively. The cosine similarity $s_{ij} = \bar{x}_i^T \bar{x}_j / \|\bar{x}_i\| \cdot \|\bar{x}_j\|$ of the feature $x_i \in \mathbb{R}^{T \times d}$ and $x_j \in \mathbb{R}^{T \times d}$ output of each layer of the feature encoder. $\bar{x}_i$ represents the average in the time dimension.

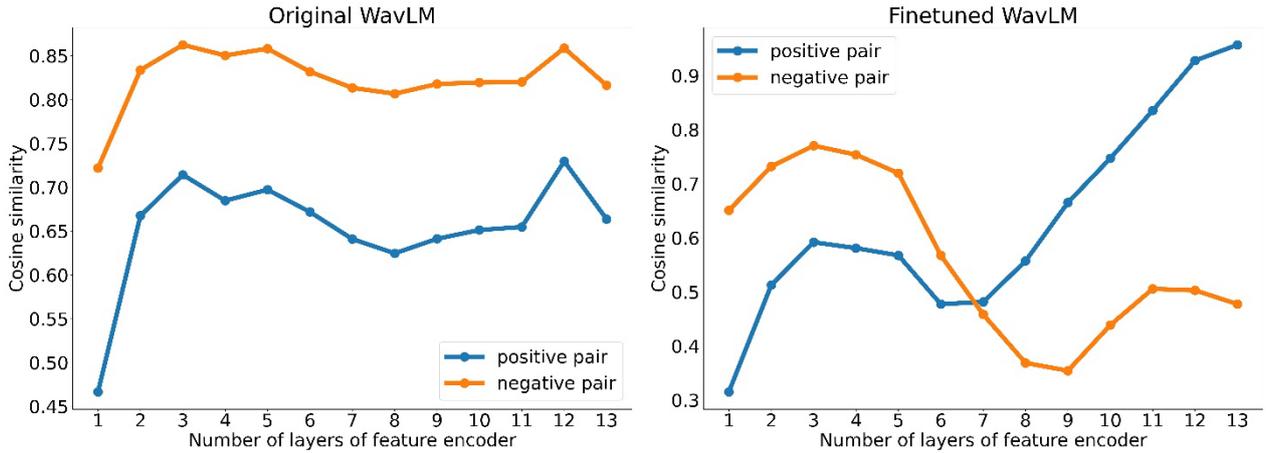

Figure 4 Feature similarity of the feature encoder output positive and negative pairs

It can be seen from the figure that for the original WavLM, the similarity of the negative pair is even higher than that of the positive pair, rising first and then leveling off as the number of feature encoder layers increases. After the fine-tuning of the first stage of WavLM, when the number of layers of the feature encoder is more than 7 layers, the similarity of the positive pair exceeds the similarity of the negative pair, and by the last layer tends to 1, while the similarity of the negative pair decreases and fluctuates around 0.4.

## 5 Conclusion

In this paper, we introduce a two-stage fine-tuning method based on supervised contrast learning. The first stage fine-tuning is performed on multiple SER datasets, and then the second stage fine-tuning is performed on a single dataset to train the classifier, and finally the optimal results can be obtained on multiple datasets simultaneously. However, the model parameters obtained by this method are relatively large, and the cost of deployment in real application scenarios is relatively high. In the future, the model distillation method can be considered to train the model with smaller parameters. In addition, fine-tuning more datasets to allow the model to learn more can enhance the model's generalization ability. In the future, human-computer interaction may use multi-modal information for emotion analysis, and use contrast learning for multi-modal alignment, which can improve the intelligence, emotional understanding ability and user experience of human-computer interaction system.